\documentclass[aps,preprint]{revtex4}%
\usepackage{amsfonts}
\usepackage{amsmath}
\usepackage{amssymb}
\usepackage{graphicx}%
\begin{document}
\title{Statistical entropy of BTZ black holes in topologically massive gravity}
\author{Baocheng Zhang}
\email{zhangbc@wipm.ac.cn}
\affiliation{State Key Laboratory of Magnetic Resonances and Atomic and Molecular Physics,
Wuhan Institute of Physics and Mathematics, Chinese Academy of Sciences, Wuhan
430071, People's Republic of China}
\keywords{Cardy formula; statistical interpretation; logarithmic correction}
\pacs{04.70.Dy, 04.50.+h}

\begin{abstract}
The entropy of a Ba\~{n}ados, Teitelboim, and Zanelli black hole in
topologically massive gravity had been given with the form inconsistent with
the Bekenstein-Hawking entropy. In the paper, we provide a consistent
statistical interpretation for the entropy, and confirm it to be thermodynamic
entropy by preserving the first and second laws. As a novel extension, the
logarithmic correction to the black hole entropy is also discussed.

\end{abstract}
\maketitle

\section{Introduction}

The black-hole solution of three-dimensional (3D) Einstein gravity with a
negative cosmological constant was found by Ba\~{n}ados, Teitelboim and
Zanelli (BTZ) \cite{btz92}, and its metric takes the form,
\begin{equation}
ds^{2}=-N^{2}dt^{2}+N^{-2}dr^{2}+r^{2}\left(  N^{\phi}dt+d\phi\right)
^{2},\label{btz}%
\end{equation}
where $\phi$ is an angle with the period $2\pi$ as the identification of the
black-hole spacetime. The functions $N^{2}$ and $N^{\phi}$ are
\begin{equation}
N^{2}=-8Gm+\frac{r^{2}}{\ell^{2}}+\frac{16G^{2}j^{2}}{r^{2}},N^{\phi}%
=\frac{4Gj}{r^{2}},
\end{equation}
where $\ell$ is the AdS radius and $G$ the 3D Newton constant. Its Killing
horizons are found by setting $N^{2}=0$; this gives
\begin{equation}
r_{\pm}=\sqrt{2G\ell\left(  \ell m+j\right)  }\pm\sqrt{2G\ell\left(  \ell
m-j\right)  },
\end{equation}
We may assume without loss of generality that $j\geqslant0$ and assume that
$\ell m\geqslant j$, to ensure the existence of an event horizon at $r=r_{+}$.

The recognition of the parameters ($m,j$) by conserved quantities of a
black-hole is significant for the black-hole thermodynamics \cite{bch73}, in
particular for the black hole entropy. For example, in the normal 3D Einstein
gravity, the parameters ($m,j$) can be interpreted as the mass $M$ and the
angular momentum $J$ of the black hole in units for which $\hbar=1$; i.e.,
\begin{equation}
M=m,J=j.\label{nema}%
\end{equation}
Here the entropy of the BTZ black hole, that is 1/4 of the length of the event
horizon in Planck units, agrees with the Bekenstein-Hawking formula
\cite{jdb73,swh75}. In particular, there is some understanding of the
microscopic degrees of freedom \cite{as98,bss98} responsible for this entropy
in terms of a boundary conformal field theory (CFT) and an application of
Cardy formula \cite{jlc86}, for which the central charge is known in the
semiclassical limit in which it is large \cite{bh86}.

In the exotic 3D Einstein gravity \cite{ew88,tz13}, however, the parameters
($m,j$) can be reinterpreted as the following forms,%
\begin{equation}
M_{E}=\frac{j}{\ell},J_{E}=\ell m,\label{eema}%
\end{equation}
which have the reversed roles for the mass and angular momentum, compared with
the case of the normal Einstein gravity. Thus in order to preserve the
thermodynamic laws, an entropy, proportional to the length of the inner
horizon instead of the event horizon, has to be distributed to the black hole
with the same metric (\ref{btz}) but with Eq. (\ref{eema}) for the
interpretation of the parameters. Usually, we called such black holes
\textquotedblleft exotic\textquotedblright. Recently, it was shown \cite{tz13}
that a statistical interpretation still exists in terms of a boundary CFT but
with the use of a modified Cardy formula.

It is well known that both the normal and exotic\textit{ }3D Einstein gravity
theories propagate no physical modes (both of them have no local degrees of
freedom), and the first successful attempt to generalize the 3D gravity theory
to propagate the gravitons is the topologically massive gravity (TMG)
\cite{djt82} which is gotten by incorporating the local Chern-Simons (LCS)
term into the normal 3D Einstein gravity. It is interesting to note that the
BTZ metric (\ref{btz}) also solves the field equations of TMG, but the
parameters ($m,j$) are related to the mass and angular momentum with the
forms,%
\begin{equation}
M_{T}=m+\frac{1}{\mu\ell^{2}}j,J_{T}=j+\frac{1}{\mu}m,\label{tmma}%
\end{equation}
where $\mu$ is the coupling parameter in TMG to modulate the LCS term of the
theory. The forms (\ref{tmma}) constrain the black hole entropy by fixing it
exclusively with the expression, $S_{T}$ $=\frac{\pi r_{+}}{2G}+\frac{1}%
{\mu\ell}\frac{\pi r_{-}}{2G}$, consistent with the thermodynamical first law
\cite{tz13}, but not consistent with the Bekenstein-Hawking formula since a
term about the length of inner horizon is included in the entropy expression.
The same entropy $S_{T}$ was also obtained by different methods
\cite{kl06,sns06,ss06,hhkt08,yt07}. However, whether such entropy still has a
statistical interpretation remains unclear. In particular, some recent attempt
\cite{mp08} required to modify the form of the entropy $S_{T}$ for some ranges
of the coupling parameter, which will lead to the inconsistency with the
thermodynamic first law. On the other hand, even though there is a statistical
interpretation for the entropy $S_{T}$, it is still unclear what will provide
the interpretation for it including all the ranges of the coupling parameter
$\mu$. Maybe there is a speculation or an expectation that the Cardy formula
plus modified Cardy formula used for the case of an exotic black hole
\cite{tz13} might provide the required statistical interpretation. But whether
this is true still needs to be investigated, which is the main purpose of the
paper. Moreover, for the case of TMG, not all the properties of the BTZ black
hole entropy can derive from the understanding for the cases of normal and
exotic black-hole entropies, as seen by a recent revealed property \cite{sd12}
related to TMG, that is, the product of the areas of the inner and outer
horizons is dependent on the mass. This also stimulates us to investigate the
statistical interpretation for the entropy $S_{T}$ and see if there is any
novel results, compared with the cases of normal and exotic gravity theories.

In the paper, we will first investigate the Cardy formula, in particular the
modified formula, with a fundamental method in CFT without recourse to the
thermodynamic relation $E=-\partial\ln Z/\partial\beta$ ($\beta$ is the
reciprocal of the temperature) as was done earlier \cite{tz13}. Moreover, in
our investigation of Cardy formula, the leading logarithmic correction
\cite{sc98,sc00,gks01,ljv11} to the entropy of a BTZ black hole appeared, and
this correction was usually considered as universal when calculating the
black-hole entropy using quantum theories of gravity (see Ref. \cite{sns11}
for general discussion about this). So in the paper, as an interesting
extension, we will also investigate whether there are any influence of the LCS
term on the logarithmic correction of BTZ black-hole entropy, since the
leading term of the entropy has changed due to the presence of the LCS term.

The structure of the paper is as follows. We will first revisit the
calculation of the Cardy formula and especially pay attention to the case with
a negative central charge in the second section. In the third section we will
calculate the BTZ black-hole entropy in TMG using the general Cardy formulas
and discuss the thermodynamical laws. The fourth section investigates the
influence of LCS term on the logarithmic correction of the BTZ black-hole
entropy. Finally, we discuss and summarize our results in the fifth section.

\section{Cardy formula}

As stated above, an interpretation of a 3D BTZ black-hole entropy has to be
made by the Cardy formula in the dual CFT which is the quantum counterpart of
the asymptotic symmetry of asymptotically AdS space. In this section, we will
focus on the investigation of modified Cardy formulas along the line of Ref.
\cite{sc98,sc00}.

Start with a two-dimensional CFT, in which the central charge $c$ can be
identified by the canonical analysis of the bulk theory according to the Brown
and Henneaux \cite{bh86}. Here we will not review the process of obtaining the
central charge while focus on the calculation of the Cardy formula to get the
BTZ black-hole entropy. The Virosoro algebra in the CFT are written as%
\begin{align}
\left[  \hat{L}_{m}^{+},\hat{L}_{n}^{+}\right]   &  =\left(  m-n\right)
\hat{L}_{m+n}^{+}+\frac{c_{L}}{12}m\left(  m^{2}-1\right)  \delta
_{m+n,0}\nonumber\\
\left[  \hat{L}_{m}^{-},\hat{L}_{n}^{-}\right]   &  =\left(  m-n\right)
\hat{L}_{m+n}^{-}+\frac{c_{R}}{12}m\left(  m^{2}-1\right)  \delta
_{m+n,0}\nonumber\\
\left[  \hat{L}_{m}^{+},\hat{L}_{n}^{-}\right]   &  =0.
\end{align}
From the Cardy calculation \cite{jlc86}, we have the quantity%
\begin{equation}
Z_{0}\left(  \tau,\bar{\tau}\right)  =Tre^{2\pi i\left(  \hat{L}_{0}^{+}%
-\frac{c_{L}}{24}\right)  \tau}e^{-2\pi i\left(  \hat{L}_{0}^{-}-\frac{c_{R}%
}{24}\right)  \bar{\tau}}%
\end{equation}
modular invariant in the transformation $\tau\rightarrow-\frac{1}{\tau}$. Then
the partition function on a torus of modulus $\tau$ is defined as%
\begin{equation}
Z\left(  \tau,\bar{\tau}\right)  =Tre^{2\pi i\tau\hat{L}_{0}^{+}}e^{-2\pi
i\bar{\tau}\hat{L}_{0}^{-}}=\sum\rho\left(  L_{0}^{+},L_{0}^{-}\right)
e^{2\pi i\tau L_{0}^{+}}e^{-2\pi i\bar{\tau}L_{0}^{-}},\label{pf}%
\end{equation}
where the signs $L_{0}^{+},L_{0}^{-}$ without hats mean the eigenvalues of
their corresponding operator matrix $\hat{L}_{0}^{+}$, $\hat{L}_{0}^{-}$. In
the Ref. \cite{sc00}, the author considered $\tau$ and $\bar{\tau}$ as
independent variables and then calculated the density-of-state for the left
movers with the positive central charge, i.e., $\rho\left(  L_{0}^{+}\right)
\approx\left(  \frac{c_{L}}{96\left(  L_{0}^{+}\right)  ^{3}}\right)
^{\frac{1}{4}}\exp\left[  2\pi\sqrt{\frac{1}{6}c_{L}L_{0}^{+}}\right]  $. If
the central charge for the right mover is still positive, the same process led
to the Cardy formula, by taking the logarithm of the exponential term of the
density of states; that is
\begin{equation}
S=2\pi\left(  \sqrt{\frac{1}{6}c_{L}L_{0}^{+}}+\sqrt{\frac{1}{6}c_{R}L_{0}%
^{-}}\right)  .
\end{equation}

Here we calculate the case with the negative central charge and assume without
loss of generality that $c_{R}<0$ and $L_{0}^{-}\leqslant0$. The density of
states can be expressed as
\begin{equation}
\rho\left(  L_{0}^{-}\right)  =\int d\tau e^{2\pi i\bar{\tau}L_{0}^{-}%
}Z\left(  \bar{\tau}\right)  .
\end{equation}

Then, according to the relation $Z\left(  \bar{\tau}\right)  =e^{-\frac{2\pi
ic_{R}}{24}\bar{\tau}}Z_{0}\left(  \bar{\tau}\right)  $ and the modular
invariance of $Z_{0}\left(  \bar{\tau}\right)  $, the density of states
becomes
\begin{equation}
\rho\left(  L_{0}^{-}\right)  =\int d\bar{\tau}e^{2\pi i\bar{\tau}L_{0}^{-}%
}e^{-\frac{2\pi ic_{R}}{24}\bar{\tau}}e^{-\frac{2\pi ic_{R}}{24}\frac{1}%
{\bar{\tau}}}Z\left(  -\frac{1}{\bar{\tau}}\right)  .
\end{equation}

In the following we will use the saddle point approximation to calculate the
integral. Define $f\left(  \bar{\tau}\right)  =2\pi i\bar{\tau}L_{0}^{-}%
-\frac{2\pi ic_{R}}{24}\bar{\tau}-\frac{2\pi ic_{R}}{24}\frac{1}{\bar{\tau}}$,
and at the extremum of $f\left(  \bar{\tau}\right)  $, $Z\left(  -\frac
{1}{\bar{\tau}}\right)  $ varies slowly, which had been checked in Ref.
\cite{sc98}, and thus we have%
\begin{equation}
\rho\left(  L_{0}^{-}\right)  \approx\int d\bar{\tau}e^{f\left(  \bar{\tau
}\right)  }\approx\left(  -\frac{2\pi}{\frac{d^{2}f}{d\bar{\tau}^{2}}%
|_{\bar{\tau}_{0}}}\right)  ^{\frac{1}{2}}e^{f\left(  \bar{\tau}_{0}\right)
},\label{ex}%
\end{equation}
if $\bar{\tau}_{0}$ is a maximum and $\frac{d^{2}f}{d\bar{\tau}^{2}}%
|_{\bar{\tau}_{0}}$ is negative. As seen, in an approximate extension of the
function $f\left(  \bar{\tau}\right)  \simeq f\left(  \bar{\tau}_{0}\right)
+\frac{\left(  \bar{\tau}-\bar{\tau}_{0}\right)  ^{2}}{2}\frac{d^{2}f}%
{d\bar{\tau}^{2}}|_{\bar{\tau}_{0}}$ at its extremum, the right-hand side of
(\ref{ex}) is explicitly computed by recognizing that the kernel of the
integral is the same as the kernel of a normal density with mean $\bar{\tau
}_{0}$ and variance $-\frac{1}{\frac{d^{2}f}{d\bar{\tau}^{2}}|_{\bar{\tau}%
_{0}}}$. So the problem transfers into calculating the extremum of the
$f\left(  \bar{\tau}\right)  $,%
\begin{equation}
\frac{df}{d\bar{\tau}}=2\pi iL_{0}^{-}-\frac{2\pi ic_{R}}{24}+\frac{2\pi
ic_{R}}{24}\frac{1}{\bar{\tau}^{2}}=0,\label{ex1}%
\end{equation}
and for $\left\vert L_{0}^{-}\right\vert \gg\left\vert c_{R}\right\vert $, we
obtain
\begin{equation}
\bar{\tau}_{0}=-i\sqrt{\frac{c_{R}}{24L_{0}^{-}}}.\label{va}%
\end{equation}
Then substituting the result of the Eq. (\ref{va}) into the definition of $f$
function, we have%
\begin{align}
f\left(  \bar{\tau}_{0}\right)   &  =2\pi L_{0}^{-}\sqrt{\frac{c_{R}}%
{24L_{0}^{-}}}+\frac{2\pi c_{R}}{24}\sqrt{\frac{c_{R}}{24L_{0}^{-}}}%
+\frac{2\pi c_{R}}{24}\sqrt{\frac{24L_{0}^{-}}{c_{R}}}\nonumber\\
&  \approx2\pi\left(  L_{0}^{-}\sqrt{\frac{c_{R}}{24L_{0}^{-}}}+\frac{c_{R}%
}{24}\sqrt{\frac{24L_{0}^{-}}{c_{R}}}\right)  \nonumber\\
&  \approx2\pi\left(  -\sqrt{\frac{\left\vert L_{0}^{-}\right\vert ^{2}c_{R}%
}{24L_{0}^{-}}}-\sqrt{\frac{\left\vert c_{R}\right\vert ^{2}L_{0}^{-}}%
{24c_{R}}}\right)  \nonumber\\
&  \approx-2\pi\sqrt{\frac{1}{6}c_{R}L_{0}^{-}\text{\ }},
\end{align}
where the negative sign in the third line is due to $c_{R}<0,L_{0}%
^{-}\leqslant0$ \cite{acj12}. Then using the result (\ref{ex}), we have%
\begin{equation}
\rho\left(  L_{0}^{-}\right)  \approx\left(  \frac{c_{R}}{96\left(  L_{0}%
^{-}\right)  ^{3}}\right)  ^{\frac{1}{4}}\exp\left[  -2\pi\sqrt{\frac{1}%
{6}c_{R}L_{0}^{-}}\right]  .
\end{equation}

So if $c_{L}>0$ and $L_{0}^{+}\geqslant0$, we get the complete density of
states as
\begin{equation}
\rho\left(  L_{0}^{+},L_{0}^{-}\right)  \approx\left(  \frac{c_{L}c_{R}%
}{96^{2}\left(  L_{0}^{+}L_{0}^{-}\right)  ^{3}}\right)  ^{\frac{1}{4}}%
\exp\left[  2\pi\left(  \sqrt{\frac{1}{6}c_{L}L_{0}^{+}}-\sqrt{\frac{1}%
{6}c_{R}L_{0}^{-}}\right)  \right]  ,\label{sdcf}%
\end{equation}
the exponential term of which leads to our earlier modified Cardy formula,%
\begin{equation}
S_{m}=2\pi\left(  \sqrt{\frac{1}{6}c_{L}L_{0}^{+}}-\sqrt{\frac{1}{6}c_{R}%
L_{0}^{-}}\right)  .\label{cf}%
\end{equation}
Using this formula, a statistical interpretation of the entropy of exotic BTZ
black holes is obtained in the dual CFT \cite{tz13}. In the calculation here,
we follow the presumption of Refs. \cite{sc98,sc00} and regard the calculation
about the part of $\tau$ and $\bar{\tau}$ independently, that is the
holomorphic factorization of the left and right sectors stated in Ref.
\cite{mw10}. Thus our results will not be plagued by the nonunitarity implied
by the negative central charge. Moreover, mathematically the saddle
approximation and the modular invariance required in the calculation above do
not have the direct or intrinsic relation with the unitarity of the theory.

In particular, we can extend the calculation above and obtain a general
formula in dual CFT for the purpose of explaining the entropy of BTZ black
holes statistically,%

\begin{equation}
S=\left\{
\begin{array}
[c]{c}%
2\pi\left(  \sqrt{\frac{1}{6}c_{L}L_{0}^{+}}+\sqrt{\frac{1}{6}c_{R}L_{0}^{-}%
}\right)  ,\text{ \ \ \ \ \ \ }c_{L},c_{R},L_{0}^{+},L_{0}^{-}\geqslant0\\
2\pi\left(  \sqrt{\frac{1}{6}c_{L}L_{0}^{+}}-\sqrt{\frac{1}{6}c_{R}L_{0}^{-}%
}\right)  ,\text{ \ }c_{L},L_{0}^{+}\geqslant0,c_{R},L_{0}^{-}\leqslant0\\
2\pi\left(  -\sqrt{\frac{1}{6}c_{L}L_{0}^{+}}+\sqrt{\frac{1}{6}c_{R}L_{0}^{-}%
}\right)  ,c_{L},L_{0}^{+}\leqslant0,c_{R},L_{0}^{-}\geqslant0\\
2\pi\left(  -\sqrt{\frac{1}{6}c_{L}L_{0}^{+}}-\sqrt{\frac{1}{6}c_{R}L_{0}^{-}%
}\right)  ,\text{ \ \ \ \ }c_{L},c_{R},L_{0}^{+},L_{0}^{-}\leqslant0
\end{array}
\right.  .\label{cfg}%
\end{equation}
However, whether the last two formulas have an application to the statistical
interpretation of the entropy of a 3D black hole is not clear up to now. In
the next section, we will fill the implicit use of the third formula by
finding a corresponding example in TMG.

\section{Statistical entropy of BTZ black holes in TMG}

Start with the gravitational action of TMG \cite{djt82},%
\begin{equation}
I_{T}=\frac{1}{16\pi G}%
{\displaystyle\int}
d^{3}x\sqrt{-g}\left(  R-2\Lambda+\frac{1}{\mu}%
\mathcal{L}%
_{CS}\right)  ,
\end{equation}
where $\Lambda=-1/\ell^{2}$ is the negative cosmological constant and $\mu$ is
the coupling parameter. It is easily seen that the TMG consists of normal
Einstein gravity and LCS term, i.e. $%
\mathcal{L}%
_{CS}=\frac{1}{2}\epsilon^{\mu\nu\rho}\Gamma_{\mu\beta}^{\alpha}\left(
\partial_{\nu}\Gamma_{\alpha\rho}^{\beta}+\frac{2}{3}\Gamma_{\nu\gamma}%
^{\beta}\Gamma_{\rho\alpha}^{\gamma}\right)  $ where $\Gamma$ is the
Christoffel symbols. And the BTZ black-hole solution (\ref{btz}) satisfies the
equations of motion of TMG,%
\begin{equation}
G_{\mu\nu}+\frac{1}{\mu}C_{\mu\nu}=0,
\end{equation}
where $G_{\mu\nu}=R_{\mu\nu}-\frac{1}{2}g_{\mu\nu}R$ is the 3D Einstein tensor
and $C_{\mu\nu}=\epsilon_{\mu}{}^{\tau\rho}\nabla_{\tau}\left(  R_{\rho\nu
}-\frac{1}{4}g_{\rho\upsilon}R\right)  $ is the Cotton tensor. But the mass
and the angular momentum of the black hole have the forms presented in Eq.
(\ref{tmma}), and in terms of the mass and angular momentum, the horizons
change to
\begin{equation}
r_{\pm}=\sqrt{2G\ell}\left(  a\sqrt{\left\vert M_{T}\ell+J_{T}\right\vert }\pm
b\sqrt{\left\vert M_{T}\ell-J_{T}\right\vert }\right)  \label{tmge}%
\end{equation}
where $a=\frac{1}{\sqrt{\left\vert 1+\frac{1}{\mu\ell}\right\vert }}$,
$b=\frac{1}{\sqrt{\left\vert 1-\frac{1}{\mu\ell}\right\vert }}$. Moreover,
$M_{T}\ell-J_{T}=\left(  1-\frac{1}{\mu\ell}\right)  \left(  m\ell-j\right)  $
and $M_{T}\ell+J_{T}=\left(  1+\frac{1}{\mu\ell}\right)  \left(
m\ell+j\right)  $, and their signs depend on the value of the coupling parameter.

The central charges had been calculated in Ref. \cite{hhkt08}, using the
Brown-Henneaux's canonical approach, i.e., $c_{L}=\left(  1+\frac{1}{\mu\ell
}\right)  \frac{3\ell}{2G},c_{R}=\left(  1-\frac{1}{\mu\ell}\right)
\frac{3\ell}{2G}$, and the zero modes $L_{0}^{+}$ and $L_{0}^{-}$ of Virasoro
algebras for the left and right movers are $L_{0}^{+}=\frac{1}{2}\left(
M_{T}\ell+J_{T}\right)  =\frac{1}{2}\left(  1+\frac{1}{\mu\ell}\right)
\left(  \ell m+j\right)  $ and $L_{0}^{-}=\frac{1}{2}\left(  M_{T}\ell
-J_{T}\right)  =\frac{1}{2}\left(  1-\frac{1}{\mu\ell}\right)  \left(  \ell
m-j\right)  $. Then, according to our discussion about Cardy formula in the
last section, we will calculate the entropy of BTZ black holes within the
different parameter ranges.

(a) when $\mu>0$ and $\mu\ell>1$, all central charges and the eigenvalues of
the zero modes are positive, so its entropy is calculated as%
\begin{align}
S_{T} &  =2\pi\left(  \sqrt{\frac{1}{6}c_{L}L_{0}^{+}}+\sqrt{\frac{1}{6}%
c_{R}L_{0}^{-}}\right)  \nonumber\\
&  =\pi\sqrt{\frac{\ell}{2G}}\left(  \left(  1+\frac{1}{\mu\ell}\right)
\sqrt{\ell m+j}+\left(  1-\frac{1}{\mu\ell}\right)  \sqrt{\ell m-j}\right)
\nonumber\\
&  =\frac{\pi r_{+}}{2G}+\frac{1}{\mu\ell}\frac{\pi r_{-}}{2G}.
\end{align}
This entropy is positive. In the case an lower bound of the mass is found,
$M_{T}\ell-J_{T}\geqslant0$ or $M_{T}\geqslant\frac{J_{T}}{\ell}$ which is
similar to the situation of a BTZ black hole in the normal Einstein gravity.
In particular, it is noted that the case is just that discussed in the Ref.
\cite{hhkt08}. Moreover, for the case $\mu<0$ and $\mu\ell<-1$, all central
charges and the eigenvalues of the zero modes are also positive, so its
entropy is still $S_{T}=\frac{\pi r_{+}}{2G}+\frac{1}{\mu\ell}\frac{\pi r_{-}%
}{2G}$. In particular, when $\mu\ell=\pm1$, one of the central charges
disappears, which leads to the chiral gravity \cite{lss08}, but their BTZ
black-hole entropies still conform to the above calculation, i.e.,
$S_{T}=\frac{\pi r_{+}}{2G}+\frac{1}{\mu\ell}\frac{\pi r_{-}}{2G}=\frac{\pi
r_{+}}{2G}\pm\frac{\pi r_{-}}{2G}$.

(b) when $\mu>0$ and $0<\mu\ell<1$, we have $c_{R}<0$ and $L_{0}^{-}<0$, so
its entropy should be calculated according to%
\begin{align}
S_{T} &  =2\pi\left(  \sqrt{\frac{1}{6}c_{L}L_{0}^{+}}-\sqrt{\frac{1}{6}%
c_{R}L_{0}^{-}}\right)  \nonumber\\
&  =\pi\sqrt{\frac{\ell}{2G}}\left(  \left(  1+\frac{1}{\mu\ell}\right)
\sqrt{\ell m+j}-\left(  \frac{1}{\mu\ell}-1\right)  \sqrt{\ell m-j}\right)
\nonumber\\
&  =\frac{\pi r_{+}}{2G}+\frac{1}{\mu\ell}\frac{\pi r_{-}}{2G}.
\end{align}
This entropy is also positive. In the case an upper bound of the mass is
found, $M_{T}\ell-J_{T}\leqslant0$ or $M_{T}\leqslant\frac{J_{T}}{\ell}$ which
is similar to the situation of an exotic BTZ black hole.

(c) when $\mu<0$ and $-1<\mu\ell<0$, we have $c_{L}<0$ and $L_{0}^{+}<0$, so
its entropy should be calculated according to%
\begin{align}
S_{T} &  =2\pi\left(  -\sqrt{\frac{1}{6}c_{L}L_{0}^{+}}+\sqrt{\frac{1}{6}%
c_{R}L_{0}^{-}}\right)  \nonumber\\
&  =\pi\sqrt{\frac{\ell}{2G}}\left(  -\left\vert 1+\frac{1}{\mu\ell
}\right\vert \sqrt{\ell m+j}+\left(  1-\frac{1}{\mu\ell}\right)  \sqrt{\ell
m-j}\right)  \nonumber\\
&  =\frac{\pi r_{+}}{2G}+\frac{1}{\mu\ell}\frac{\pi r_{-}}{2G}.
\end{align}
where it is noted that the positivity of the entropy requires $-1<\mu
\ell\leqslant-\frac{r_{-}}{r_{+}}$, because, for the microcanonical ensemble,
the negative entropy means the number of microstates in the CFT is less than
$1$, which is hard to accept physically. Moreover, it is interesting to note
that the excluded range could be expressed as $-\frac{r_{-}}{r_{+}}<\mu\ell<0$
or $-1<\frac{\mu}{\Omega}<0$ where $\Omega=\frac{r_{-}}{\ell r_{+}}$ is the
angular velocity of BTZ black holes, and so in order to preserve the positive
entropy, the coupling strength $\left\vert \mu\right\vert $ cannot be smaller
than the angular velocity $\Omega$. On the other hand, we note that the
excluded range $-\frac{r_{-}}{r_{+}}<\mu\ell<0$ leads to the negative mass of
BTZ black holes in TMG, i.e. $\ell M_{T}=\ell m+\frac{1}{\mu\ell}j<\ell
m-\frac{r_{+}}{r_{-}}j=\ell m-\frac{\sqrt{\ell m+j}+\sqrt{\ell m-j}}%
{\sqrt{\ell m+j}-\sqrt{\ell m-j}}j=\ell m-\frac{lm+\sqrt{\ell m+j}\sqrt{\ell
m-j}}{j}j=-\sqrt{\ell m+j}\sqrt{\ell m-j}\leq0$. But the linearized analysis
gave the classical stability of the BTZ black hole in TMG for all the values
of the coupling parameter \cite{bms10}. This is peculiar for a stable black
hole with a negative mass and so it deserves further investigation for the
stability of the BTZ black hole in the background of TMG. Moreover, if the
conclusion of Ref. \cite{bms10} is true, we could guess that the negative sign
of the entropy in the range $-\frac{r_{-}}{r_{+}}<\mu\ell<0$ might only be due
to the negative sign of the mass. Since the black hole in the range
$-\frac{r_{-}}{r_{+}}<\mu\ell<0$ could still be stable, the corresponding
number of its microscopical states might be suggested accordingly as
$e^{\left\vert S_{T}\right\vert }$, which remains to be confirmed in the
future quantum gravity theory.

Thus, based on our analysis using the Cardy formula, the counting of the
microstates gives the same form of the entropy,
\begin{equation}
S_{T}=\frac{\pi r_{+}}{2G}+\frac{1}{\mu\ell}\frac{\pi r_{-}}{2G}\label{etmg}%
\end{equation}
which is different from the suggested results in Ref. \cite{mp08} but
consistent with many other methods \cite{sns06,kl06,yt07}, and the entropy is
positive except in case (c) the more tightened condition is required.
Moreover, the cases (a)-(c) have not included the use of the last formula of
the general Cardy formula (\ref{cfg}), and so it is still unclear for its
possible function as a statistical formula for some black-hole entropies. To
some extent, this also provided an interesting motivation to find a 3D gravity
theory with both the left and right negative central charges.

Then we have to investigate whether the entropy (\ref{etmg}) obtained
statistically satisfies the laws of thermodynamics. First the entropy
satisfies the first law
\begin{equation}
dM_{T}=T_{H}dS_{T}+\Omega dJ_{T}%
\end{equation}
with the thermodynamic variables $T_{H}=\frac{r_{+}^{2}-r_{-}^{2}}{2\pi
r_{+}\ell^{2}},\Omega=\frac{r_{-}}{\ell r_{+}}$ which are geometrical in the
sense that they depend only on the location of the Killing horizons and are
model independent, i.e., independent of the specific field equations that are
solved by the BTZ metric. Remarkably, in our earlier paper \cite{tz13}, it was
pointed out that for the forms (\ref{tmma}) of the mass and the angular
momentum, only the entropy (\ref{etmg}) obtained here can preserve the first
law. Thus our calculation here confirms that the entropy obtained by a
statistical method is consistent with the requirement of black-hole
thermodynamic laws.

Once the first law is held, it is easier to confirm the second law by the
so-called method of \textquotedblleft physical process\textquotedblright%
\ \cite{jkm95}, i.e. $dS_{T}=\frac{1}{T_{H}}\left(  dM_{T}-\Omega
dJ_{T}\right)  =\frac{2}{T_{H}}dM_{T}>0$ for the process of the particle
absorption by a black hole, where the relations $\frac{dM_{T}}{dJ_{T}}%
=\frac{dM_{T}}{dS_{T}}\frac{dS_{T}}{dJ_{T}}=-\Omega$ and $S_{T}=S_{T}\left(
M_{T},J_{T}\right)  $ are used. In particular, for BTZ black holes in any 3D
gravity theory, the particle absorption generally leads to the change of the
entropy $dS_{T}=\frac{2}{T_{H}}dM_{T}$, which is similar to the situation of
the Schwarzschild black hole in four-dimensional spacetime, for which the
change of the entropy is proportional to the term $\frac{\Delta E}{M}$, where
$\Delta E$ is the energy of absorbed particles. Of course that is only valid
for the quasistationary situation, but a general analysis of the particle
absorbing process had been made in a recent work \cite{gl12}, and it is easy
to apply and extend their conclusions to the situation of TMG.

\section{Logarithmic correction}

From the above discussion, we knew that the LCS term gave a BTZ black-hole
entropy associated with the outer (event) horizon a term proportional to the
length of the inner horizon. The peculiar behavior made the author of the Ref.
\cite{sns06} speculate that the observer at infinity might see the interior of
the black hole when the LCS term was included. Thus it is interesting to
explore whether the influence of the LCS term could extend to the logarithmic
correction of the black-hole entropy associated with the outer horizon, since
the logarithmic correction is usually believed to be caused by the quantum
gravity effect \cite{km00,zcz08} in the calculation using the methods of
string theory \cite{sns98} and loop quantum gravity \cite{gm05}.

From the density of states presented in Eq. (\ref{sdcf}), one can get the
logarithmic correction of the entropy associated with the outer horizon by the
part before exponential term,%
\begin{equation}
\Delta S=\frac{1}{4}\ln\left(  \frac{c_{L}c_{R}}{96^{2}\left(  L_{0}^{+}%
L_{0}^{-}\right)  ^{3}}\right)  .
\end{equation}
For the normal 3D Einstein gravity, it had been calculated \cite{sc00,gks01}
as%
\begin{equation}
\Delta S_{G}=-\frac{3}{2}\ln\frac{\pi r_{+}}{2G}-\frac{3}{2}\ln\kappa
\ell-\frac{3}{2}\ln\frac{\pi\ell}{2G}-\ln8G\ell^{2},\label{ge}%
\end{equation}
where $\kappa=\frac{r_{+}^{2}-r_{-}^{2}}{r_{+}\ell^{2}}$ is the surface
gravity of the outer horizon, and it is also geometric and a constant for a
specific BTZ black hole. For the exotic 3D Einstein gravity, the logarithmic
correction is the same as in Eq. (\ref{ge}) up to a different constant term.

Then for TMG, the correction of the entropy is%
\begin{equation}
\Delta S_{T}=-\frac{3}{2}\ln\frac{\pi r_{+}}{2G}-\frac{3}{2}\ln\kappa
\ell-\frac{3}{2}\ln\frac{\pi\ell}{2G}-\ln8G\ell^{2}ab.\label{te}%
\end{equation}
where $a$ and $b$, the same as that presented in Eq. (\ref{tmge}), are related
to the coupling parameter $\mu$. Comparing Eqs. (\ref{ge}) and (\ref{te}), we
find that the difference exists only in the last constant term. This shows
that even in the gravity theories with the LCS term, the logarithmic
correction of the entropy associated with the outer horizon is not related to
the length of the inner horizon. Thus it hints that the influence of LCS term
on the entropy of the black hole only works for the leading order term and
cannot be extended into the deeper knowledge of the black-hole entropy,
i.e\textit{.,} logarithmic correction.

\section{Discussion and Conclusion}

In the paper, we have investigated the Cardy formula and have made the
calculation for the case with the negative central charge. Our results showed
that in the assumption of holomorphic factorization, for the same BTZ
black-hole metric form and the same asymptotic AdS condition, the Cardy
formula is \textit{not} completely the same and depends on the sign of the
central charges and zero eigenvalues of the left and right movers $\hat{L}%
_{m}^{\pm}$. This validated and extended our earlier result \cite{tz13}. We
have also applied these results to the BTZ black holes in the TMG and obtained
the same entropy as that by other different methods. Thus we provided a
statistical interpretation again for a non-Bekenstein-Hawking entropy. In
particular, different from the exotic Einstein gravity, the BTZ black-hole
entropy in TMG has to be explained statistically through different Cardy
formulas in the different coupling parameter ranges. It is also noted that the
statistical entropy of BTZ black holes in higher curvature gravity had been
calculated \cite{ss00}, and when the LCS term was added to the higher
curvature gravity, e.g., general massive gravity theory \cite{bht09}, the
calculation of the leading term by the Cardy formula still gives a
non-Bekenstein-Hawking entropy statistically. After calculating the black-hole
entropy statistically in TMG, we have also discussed the thermodynamic first
and second laws, and found that they both apply for such entropy. Thus the
obtained statistical entropy is still thermodynamic entropy, although it does
not take the Bekenstein-Hawking form.

It is noted that in all the three cases involved, i.e. the normal and exotic
3D Einstein gravity and TMG, there is an obvious difference which is reflected
on the parity of the respective bulk theories of gravity. Although it is still
unclear for the role of parity in the statistical interpretation of the
black-hole entropy, the advantage of definite parity in calculating the
black-hole entropy using the method of canonical quantization has been
discussed \cite{cv00}. Thus what role the parity of the general gravity theory
would play and how it would play in the process of calculating the
corresponding black hole entropy deserves further investigation.

The derivation of non-Bekenstein-Hawking entropy is due to the introduction of
the LCS term which might be the reason for scrambling the parity in TMG. From
our analysis, although the LCS term does not influence the thermodynamics of
the outer horizon, it indeed gives a peculiar term which is proportional to
the length of the inner horizon in the expression of the black-hole entropy of
the outer horizon. We have investigated whether the LCS term influenced the
term of logarithmic corrections to the black-hole entropy associated with the
outer horizon, and found the answer to be no, any influences existed up to a
constant term. We concluded that the influence of the LCS term will not extend
to the deeper level of the black-hole entropy at least in 3D spacetime, which
will be significant for understanding exactly the role of LCS term in the
general 3D gravity or quantum gravity theories.

\section{Acknowledgements}

The author would like to thank Prof. P. K. Townsend for his helpful
suggestions and discussions and also acknowledges the support from Grant Nos.
11104324, 11374330 of the National Natural Science Foundation of China.


\begin{thebibliography}{99}                                                                                               %


\bibitem {btz92}M. Ba\~{n}ados, C. Teitelboim, and J. Zanelli, Phys. Rev.
Lett. \textbf{69} 1849 (1992).

\bibitem {bch73}J. M. bardeen, B. Carter, and S. W. Hawking, Commun. Math.
Phys. \textbf{31}, 161 (1973).

\bibitem {jdb73}J. D. Bekenstein, Phys. Rev. D \textbf{7}, 2333 (1973).

\bibitem {swh75}S. W. Hawking, Commun. Math. Phys. \textbf{43}, 199 (1975);
\textbf{46}, 206 (1976).

\bibitem {as98}A. Strominger, J. High Energy Phys. \textbf{02}, 009 (1998).

\bibitem {bss98}D. Birmingham, I. Sachs, and S. Sen, Phys. Lett. B
\textbf{424}, 275 (1998).

\bibitem {jlc86}J. L. Cardy, Nucl. Phys. B \textbf{270}, 186 (1986).

\bibitem {bh86}J. D. Brown and M. Henneaux, Commun. Math. Phys. \textbf{104},
207 (1986).

\bibitem {ew88}E. Witten, Nucl. Phys. B \textbf{311}, 46 (1988).

\bibitem {tz13}P. K. Townsend and B. Zhang, Phys. Rev. Lett. \textbf{110},
241302 (2013).

\bibitem {djt82}S. Deser, R. Jackiw, and S. Templeton, Ann. Phys. (N.Y.)
\textbf{140}, 372 (1982); \textbf{185}, 406 (1988); \textbf{185}, 406 (1988);
\textbf{281}, 409 (2000).

\bibitem {kl06}P. Kraus and F. Larsen, J. High Energy Phys. \textbf{01} 022 (2006).

\bibitem {sns06}S. N. Solodukhin, Phys. Rev. D \textbf{74}, 024015 (2006).

\bibitem {ss06}B. Sahoo and A. Sen, J. High Energy Phys. \textbf{07} 008 (2006).

\bibitem {hhkt08}K. Hotta, Y. Hyakutake, T. Kubota, and H. Tanida, J. High
Energy Phys. \textbf{07} (2008) 066.

\bibitem {yt07}Y. Tachikawa, Class. Quantum Grav. \textbf{24} 737 (2007).

\bibitem {mp08}Mu-In Park, Phys. Rev. D \textbf{77}, 026011 (2008).

\bibitem {sd12}S. Detournay, Phys. Rev. Lett. \textbf{109} 031101 (2012).

\bibitem {sc98}S. Carlip, Classical Quantum Gravity \textbf{15}, 3609 (1998).

\bibitem {sc00}S. Carlip, Classical Quantum Gravity \textbf{17}, 4175 (2000).

\bibitem {gks01}T. R. Govindarajan, R. K. Kaul, and V. Suneeta, Classical
Quantum Gravity \textbf{18}, 2877 (2001).

\bibitem {ljv11}F. Lorana, M. M. Sheikh-Jabbarib, and M. Vincon, J. High
Energy Phys. \textbf{01}, 110 (2011).

\bibitem {sns11}S. N. Solodukhin, Living Rev. Relativity \textbf{14}, 8 (2011).

\bibitem {acj12}H. Afshar, B. Cvetkovic, S. Ertl, D. Grumiller, and N.
Johansson, Phys. Rev. D \textbf{85} 064033 (2012).

\bibitem {mw10}A. Maloney and E. Witten, J. High Energy Phys. \textbf{02}, 029 (2010).

\bibitem {lss08}W. Li, W. Song and A. Strominger, J. High Energy Phys.
\textbf{04}, 082 (2008).

\bibitem {bms10}D. Birmingham, S. Mokhtari, and I. Sachs, Phys. Rev. D
\textbf{82}, 124059 (2010).

\bibitem {jkm95}T. Jacobson, G. Kang, and R. C. Myers, Phys. Rev. D
\textbf{52}, 3518 (1995).

\bibitem {gl12}B. Gwak and B.-H. Lee, Class. Quantum Grav. \textbf{29}, 175011 (2012).

\bibitem {km00}R.K. Kaul, P. Majumdar, Phys. Rev. Lett. \textbf{84} 5255 (2000).

\bibitem {zcz08}B. Zhang, Q. Y. Cai, and M. S. Zhan, Phys. Lett. B
\textbf{665} 260 (2008).

\bibitem {sns98}S. N. Solodukhin, Phys. Rev. D \textbf{57} 2410 (1998).

\bibitem {gm05}A. Ghosh, P. Mitra, Phys. Rev. D \textbf{71} 027502 (2005).

\bibitem {ss00}H. Saida and J. Soda, Phys. Lett. B \textbf{471}, 358 (2000).

\bibitem {bht09}E. A. Bergshoeff, O. Hohm, and P. K. Townsend, Phys. Rev.
Lett. \textbf{102}, 201301 (2009).

\bibitem {cv00}C. Vaz, Phys. Rev. D \textbf{61}, 064017 (2000).
\end{thebibliography}
\end{document}